# Quantization Conditions, 1900–1927

Anthony Duncan and Michel Janssen

March 9, 2020

## 1 Overview

We trace the evolution of quantization conditions from Max Planck's introduction of a new fundamental constant ($h$) in his treatment of blackbody radiation in 1900 to Werner Heisenberg's interpretation of the commutation relations of modern quantum mechanics in terms of his uncertainty principle in 1927.

In the most general sense, quantum conditions are relations between classical theory and quantum theory that enable us to construct a quantum theory from a classical theory. We can distinguish two stages in the use of such conditions. In the first stage, the idea was to take classical mechanics and modify it with an additional quantum structure. This was done by cutting up classical phase space. This idea first arose in the period 1900–1910 in the context of new theories for black-body radiation and specific heats that involved the statistics of large collections of simple harmonic oscillators. In this context, the structure added to classical phase space was used to select equiprobable states in phase space. With the arrival of Bohr's model of the atom in 1913, the main focus in the development of quantum theory shifted from the statistics of large numbers of oscillators or modes of the electromagnetic field to the detailed structure of individual atoms and molecules and the spectra they produce. In that context, the additional structure of phase space was used to select a discrete subset of classically possible motions. In the second stage, after the transition to modern quantum quantum mechanics in 1925–1926, quantum theory was completely divorced from its classical substratum and quantum conditions became a means of controlling the symbolic translation of classical relations into relations between quantum-theoretical quantities, represented by matrices or operators and no longer referring to orbits in classical phase space.[1]

More specifically, the development we trace in this essay can be summarized as follows. In late 1900 and early 1901, Planck used discrete energy units $\varepsilon = h\nu$ in his statistical treatment of radiating charged harmonic oscillators with resonance frequency $\nu$. However, he still allowed the energy of these oscillators to take on the full continuum of values. It was not until more than five years later that Albert Einstein first showed that one only arrives at the Planck law for blackbody radiation if the energy of Planck's oscillators is restricted to integral multiples of $h\nu$.

---

1. Cf. the concise account of the emergence of quantum mechanics by Darrigol (2009).

This Planck-Einstein condition was inextricably tied to a particular mechanical system, i.e., a one-dimensional simple harmonic oscillator. In his lectures on radiation theory, published in 1906, Planck suggested a more general condition for one-dimensional bound periodic systems by carving up the phase space of such systems into areas of size $h$. In late 1915, he generalized this idea to systems with several degrees of freedom (by a slicing procedure in the multidimensional phase space). By that time, Arnold Sommerfeld had independently found a procedure to quantize phase space that turned out to be equivalent to Planck's. Using this procedure, Sommerfeld was able to generalize the circular orbits of Niels Bohr's hydrogen atom (selected through quantization of the orbital angular momentum of the electron) to a larger set of Keplerian orbits of varying size and eccentricity, selected on the basis of the quantization of phase integrals such as $\oint p\,dq = nh$, where $p$ is the momentum conjugate to some generalized coordinate $q$ and the integral is to be taken over one period of the motion. Bohr's angular momentum quantum number gave the value of just one such an integral, for the angular coordinate. Quantized orbits with different eccentricities became possible once the phase integral for the radial coordinate was similarly subjected to quantization. It was soon realized by the astronomer Karl Schwarzschild that the procedures of Planck and Sommerfeld are equivalent and that they amount to treating the classical problem in action-angle variables $(p_i, q_i)$, familiar from celestial mechanics, with the action variables restricted to multiples of $h$, $J_i = \oint p_i\,dq_i = n_i h$.

The transition from the old to the new quantum theory began in 1924 with the transcription (inspired by Bohr's correspondence principle) of the classical action derivative $d/dJ$ as a discrete difference quotient, $(1/h)\Delta/\Delta n$, by Hans Kramers, John Van Vleck, Max Born, and others. This transcription procedure was critical in the development of Kramers's dispersion theory for the elastic scattering of light. It led to the introduction of complex coordinate amplitudes depending on a pair of states linked by a quantum transition. In his famous *Umdeutung* (i.e., reinterpretation) paper, Heisenberg reinterpreted these amplitudes as two-index arrays with a specific non-commutative multiplication rule. Applying the transcription procedure to the Sommerfeld phase integral itself, he arrived at a nonlinear quantization constraint on these amplitudes. He showed that this constraint is just the high-frequency limit of the Kramers dispersion formula, known as the Thomas-Kuhn sum rule. Born quickly recognized that Heisenberg's two-index arrays are nothing but matrices and that the multiplication rule is simply the rule for matrix multiplication. Rewriting Heisenberg's quantization condition in matrix language, Born and Pascual Jordan arrived at the familiar commutation relation $[p_k, q_l] \equiv p_k q_l - q_l p_k = (\hbar/\mathrm{i})\delta_{kl}$ of modern quantum mechanics (with $\hbar \equiv h/2\pi$ and $\delta_{kl}$ the Kronecker delta). Around the same time and independently of Born and Jordan, Paul Dirac showed that this commutation relation is the exact analogue of Poisson brackets in classical mechanics. The commutation relation for position and momentum was also found to be satisfied by the operators representing these quantities and acting on wave functions in the alternative form of quantum mechanics developed by Erwin Schrödinger in late 1925 and early 1926. This commutation relation represents the central locus for the injection of Planck's constant into the new quantum theory. In 1927, Heisenberg interpreted it in terms of his uncertainty principle.

In the balance of this chapter, we examine the developments sketched above in more detail.[2]

## 2 The earliest quantum conditions

In December 1900, Planck introduced the relation $\varepsilon = h\nu$ to provide a derivation of the empirically successful new formula he had first presented a couple of months earlier for the spectral distribution of the energy in blackbody radiation. Unlike Einstein and, independently, Paul Ehrenfest, Planck did not apply the relation $\varepsilon = h\nu$ to the energy of the radiation itself but to the energy of tiny charged harmonic oscillators (which he called "resonators"), spread throughout the cavity and interacting with the radiation in it. Planck used the relation $\varepsilon = h\nu$ to count the number of possible microstates of a collection of such resonators. He then inserted this number into Boltzmann's formula relating the entropy of a macrostate to the number of microstates realizing it. The formula Planck thus found for the entropy of a resonator leads directly to the law for blackbody radiation now named after him. Planck did not restrict the possible values of the resonator energy to integral multiples $nh\nu$ (where $n$ is an integer), he only assumed that energies between successive values of $n$ should be lumped together when counting microstates. It was not until six years later that Einstein finally showed that Planck's derivation only works if the resonator energies are, in fact, restricted to integral multiples of $h\nu$.[3]

---

2. For much more detailed accounts, see Jammer (1966), Mehra and Rechenberg (1982–2001), Darrigol (1992), and Duncan and Janssen (2019–2022).

3. Planck (1900a, 1900b, 1901), Einstein (1905), Ehrenfest (1906), Einstein (1906). Following the publication of Kuhn's (1978) revisionist account of the origins of quantum theory, historians have reevaluated Planck's work and its early reception. Some of the most prominent contributions to this reevaluation are (in roughly chronological order): Klein's review of Kuhn's book (Klein, Shimony, and Pinch 1979, pp. 430–434), Kuhn's (1984) response to his critics, Needell (1980, 1988), Darrigol (1992, 2000, 2001) and Gearhart (2002). This debate informed our discussion of the early history of quantum theory in Duncan and Janssen (2019–2022, Vol. 1, Chs. 2–3).

That same year, 1906, Planck published his lectures on radiation theory, in which he reworked his own discretization of resonator energy by dividing the phase space of a resonator, spanned by its position and momentum, into cells of size $h$. In 1908, he finally accepted that the energy of a resonator is quantized: a resonator is only allowed to be at the edges of the cells in phase space. In 1911, he once again changed his mind in what came to be known as "Planck's second theory". He now proposed that a resonator can absorb energy from the ambient radiation continuously but release energy only when its own energy is an integral multiple of $h\nu$ and then only in integral multiples $nh\nu$.[4]

In 1913, drawing both on the quantum theory of blackbody radiation and on the British tradition of atomic modeling, Bohr, in the first part of his famous trilogy, proposed a quantum model of the hydrogen atom. He showed that this model gives the correct formula for the Balmer lines with a value for the Rydberg constant in excellent agreement with the spectroscopic data. Bohr quantized the energy of the electron in the hydrogen atom but eventually settled on quantizing its angular momentum instead, restricting its value to integer multiples of $h/2\pi$ (or, in modern notation, $\hbar$). Although he allowed elliptical orbits, Bohr did most of his calculations for circular orbits. This did not affect his results as he only used one quantization condition. In the second and third parts of his trilogy, he used this same quantization condition for planar models of more complicated atoms and molecules.[5]

## 3  Quantization conditions in the old quantum theory

Bohr's success in accounting for the most prominent features of the hydrogen spectrum led to a shift in work on quantum theory. Instead of dealing with large collections of harmonic oscillators, physicists began to focus on individual atoms and tried to account for their spectra with or without external electric or magnetic fields on the basis of models similar to Bohr's. These efforts resulted in what, after the transition to modern quantum mechanics in the mid-1920s, came to be known as the old quantum theory. Its undisputed leader, besides Bohr in Copenhagen, was Sommerfeld in Munich. The development of the old quantum theory can be followed in successive editions of his book *Atomic Structure and Spectral Lines* (*Atombau und Spektrallinien*), which became known as the bible of atomic theory. Ehrenfest referred to its

---

4. Planck (1906), Planck to Lorentz, October 7, 1908 (Lorentz 2008, Doc. 197; for discussion of this letter, see Kuhn 1987, p. 198). For "Planck's second theory" see the second edition of his lectures on radiation theory (Planck 1913).

5. Bohr (1913). On the genesis and reception of the Bohr model, see Heilbron and Kuhn (1969) and Kragh (2012).

author (though not as a compliment) as the theory's pope.[6]

In two papers presented to the Munich Academy in December 1915 and January 1916, Sommerfeld rephrased Bohr's quantization condition in terms of Planck's phase-space quantization, with the understanding that only states at the edge of Planck's cells are allowed. Sommerfeld elaborated on these ideas in a paper published in two installments in *Annalen der Physik*. In one dimension, this phase-space quantization rule restricts the values of what were called phase integrals, the integral of the conjugate momentum $p$ of some generalized coordinate $q$ over one period of the motion, to integer multiples of $h$: $\oint p \, dq = nh$. This allowed Sommerfeld to subsume the quantized oscillators of Planck and Einstein and Bohr's model of the hydrogen atom under one quantization rule. Moreover, he generalized Bohr's model by allowing elliptical as well as circular orbits. He accomplished this by applying his phase-space quantization rule both to the radial coordinate and to the angular coordinate and their conjugate momenta. He recovered Bohr's quantum number $n$ as the sum of the two quantum numbers $n_r$ and $n_\varphi$ that he had introduced to quantize the orbits in a hydrogen atom in polar coordinates. The energy levels Bohr had identified thus correspond to multiple orbits with different combinations of Sommerfeld's radial and angular quantum numbers. This degeneracy is lifted, Sommerfeld showed, when the relativistic dependence of the mass of the electron on its velocity is taken into account. Sommerfeld thus found a formula for the fine structure of the hydrogen spectrum. This formula survives to this day even though, compared to modern quantum mechanics, Sommerfeld's quantum numbers for angular momentum are all off by 1. Even more baffling, he derived it without any knowledge of electron spin.[7]

As Sommerfeld was adapting Bohr's quantization condition to Planck's phase space ideas, Planck himself, in two presentations to the German Physical Society in November and December 1915, tried to generalize the phase-space slicing he had introduced for one-dimensional oscillators to systems of multiple degrees of freedom. In a system described by $n$ independent canonical coordinates $q_1, \ldots, q_n$ and $n$ conjugate momenta $p_1, \ldots, p_n$, the $2n$ dimensional phase-space was to be sliced into cells of equal phase-space volume $h^n$. The surfaces producing this slicing were prescribed by $n$ functions $g_i(q_j, p_k)$ (with $i, j, k = 1, \ldots, n$), subject to the quantization conditions $g_i = n_i h$ at the boundaries of the cells. Planck placed two requirements on these functions. First, during the completely classical motion of the system in between the quantum jumps characteristic of transitions between stationary states in the Bohr picture, the system should either stay between the boundaries of two cells or move along one of these boundaries. The simplest way to enforce this is the one adopted by Planck: choose phase-space functions $g_i$ that are constants of the motion. Second, the requirement that cells take up equal volumes of phase space entails a factorization of the phase-space measure: the functions $g_i$ had to be chosen so that the multidimensional volume $dq_1 \ldots dq_n dp_1 \ldots dp_n$ could be rewritten as $dg_1 \ldots dg_n$. Planck's procedure could only be implemented on a case-by-case basis—and rather awkwardly at that. Planck applied his method to a number of cases (two-dimensional oscillators, Coulomb problem, three-dimensional rigid body) but did not find the correct quantized energy levels in all cases. The basic idea, however, was correct. In the action-angle formalism, developed in the context of celestial mechanics and transferred to atomic physics a few months later by Schwarzschild, the two conditions that Planck imposed on the slicing functions $g_i$ are automatically satisfied by the action variables $J_i$ in cases where such variables exist. Had Planck been *au courant* with the action-angle formalism, he might well have recognized this and anticipated Schwarzschild's seminal work a few months later.[8]

---

6. Sommerfeld (1919). An English translation of the third edition appeared shortly after the publication of the German original (Sommerfeld 1923). Sommerfeld's contributions are discussed in Eckert (2013a, 2013b, 2013c, 2014).

7. Sommerfeld (1915a, 1915b, 1916). For discussion of the fortuitous character of Sommerfeld's derivation of the fine-structure formula, see Yourgrau and Mandelstam (1979) and Biedenharn (1983).

The Planck-Sommerfeld rule for quantizing phase integrals was found independently by William Wilson and Jun Ishiwara. It was left to Schwarzschild, however, to make the connection between phase integrals and action variables, well-known to astronomers knowledgeable about celestial mechanics and the techniques of Hamilton-Jacobi theory. In a letter of March 1916, Schwarzschild alerted Sommerfeld to this connection. Combining these techniques from celestial mechanics with Sommerfeld's quantum condition, Schwarzschild in short order derived a formula for the Stark effect in hydrogen, the splitting of its spectral lines in an external electric field. Sommerfeld's former student Paul Epstein arrived at essentially the same result at essentially the same time.[9]

What made action variables natural candidates for quantization was that they were so-called adiabatic invariants. As early as 1913, as can be gathered from a letter to Joffe of February that year, Ehrenfest had realized the importance for quantum theory of a theorem found independently by Kalman Szily, Rudolf Clausius, and Ehrenfest's teacher Ludwig Boltzmann. This theorem asserts that under slow changes of the parameters of a mechanical system undergoing periodic motion, the integral of its kinetic energy over a single period is time invariant. For a harmonic oscillator with its total energy restricted to integral multiples of $h\nu$, Ehrenfest realized, this means that the adiabatic invariant $\overline{E_{\text{kin}}}/\nu$, the ratio of its average kinetic energy and its characteristic frequency, has to be set equal to $\frac{1}{2}nh$. Around the same time and independently of Bohr, Ehrenfest used an adiabatic-invariance argument to quantize the angular momentum of diatomic molecules: $L = n\hbar$. As soon as he saw Sommerfeld's phase integral quantization, as he explained in a letter to his Munich colleague of May 1916, Ehrenfest made the connection with adiabatic invariants. He formally introduced what came to be known as the adiabatic principle, one of the pillars of the old quantum theory, in a paper published later that year, first in the Proceedings of the Amsterdam Academy and, shortly thereafter, in *Annalen der Physik*. In the latter paper, he formulated his "adiabatic hypothesis" in a particularly concise way: "Under reversible adiabatic transformation of a system, (quantum-theoretically) 'allowed' motions are always changed into 'allowed' motions." The following year, Jan Burgers, one of Ehrenfest's students in Leiden, supplied the proof that individual action variables are adiabatic invariants.[10]

---

8. Planck (1916).
9. Wilson (1915), Ishiwara (1915), Schwarzschild to Sommerfeld, March 1, 1916 (Sommerfeld 2000, Doc. 240), Schwarzschild (1916), Epstein (1916a, 1916b). On the history of action-angle variables, see Nakane (2015). On Schwarzschild alerting Sommerfeld to action-angle variables, see Eckert (2013a, 2014). For the explanation of the Stark effect in the old (and the new) quantum theory, see Duncan and Janssen (2014, 2015).

The application of the action-angle formalism by Schwarzschild and Epstein to the Stark effect was seen as a major success for the old quantum theory, on par with Sommerfeld's elucidation of the fine structure. It also illustrates, however, a fundamental problem that would become an important factor in the eventual demise the old quantum theory. Given the basic picture of atoms as miniature solar systems and the use of techniques borrowed from celestial mechanics to calculate the allowed energy levels, it was only natural to think of electrons as orbiting the nucleus on classical orbits. The Stark effect formed one example—its magnetic counterpart, the Zeeman effect, would provide a more dramatic one—where this picture turned out to be highly problematic. As Bohr, Sommerfeld, and Epstein realized, it makes a difference in which coordinates the quantum conditions are imposed. Even though the choice of coordinates does not affect the energy levels found, it does affect the shape of the orbits.[11]

## 4 The transition to quantum mechanics and the appearance of the modern commutation relations

A more serious problem for the picture of orbits arose in attempts to adapt the classical theory of optical dispersion of Hermann von Helmholtz, Hendrik Antoon Lorentz, and Paul Drude to the quantum theory of Bohr and Sommerfeld. The classical theory was developed to deal with the phenomenon of anomalous dispersion, the effect that the index of refraction decreases rather than increases with frequency in ranges around the absorption frequencies of the material under study. This phenomenon could be explained on the assumption that matter contains large numbers of charged oscillators with resonance frequencies at these absorption frequencies.[12]

---

10. Ehrenfest to Joffe, February 20, 1913, quoted and discussed by Klein (1970, p. 261), Ehrenfest (1913a; 1913b; 1916a; 1916b, the passage we quoted from this last paper can be found on p. 328), Ehrenfest to Sommerfeld, May 1916 (Sommerfeld 2000, Doc. 254; quoted in Klein 1970, p. 286), Burgers (1917a, 1917b). For further discussion of the adiabatic principle, see Navarro and Pérez (2004, 2006), Pérez (2009), and Duncan and Pérez (2016).
11. Duncan and Janssen (2014, 2015).

These oscillating charges could not simply be replaced by the orbiting electrons of the Bohr-Sommerfeld model of the atom. To recover the Balmer formula, Bohr had been forced to sever the relation between the orbital frequency of the electron circling the nucleus in the hydrogen atom and the frequency of the radiation emitted or absorbed upon quantum jumps of the electron from one orbit to another, a frequency given by the Bohr frequency condition $h\nu = |E_i - E_f|$ (where the subscripts $i$ and $f$ refer to initial and final orbit). Only in the limit of high quantum numbers do these transition frequencies merge with orbital frequencies. In 1913, Bohr had actually used the requirement that these two frequencies merge in this limit to put the quantum condition he needed to recover the Balmer formula on a more secure footing. In the limit of high quantum numbers, Bohr's quantum theory thus merged with classical electrodynamics according to which the frequencies of radiation are always (overtones of) the frequencies of the oscillations generating the radiation. Over the next few years, Bohr greatly expanded the use of analogies with classical electrodynamics to develop his own quantum theory. He eventually introduced the term "correspondence principle" to characterize this approach. Severing radiation and orbital frequencies meanwhile was widely seen as the most radical aspect of Bohr's model. It also meant that dispersion becomes anomalous at frequencies that differ sharply from the orbital frequencies of the electrons. Dispersion thus posed a serious problem for the old quantum theory.[13]

In 1921, Rudolf Ladenburg, an experimental physicist in Breslau, addressed a problem for the *classical* dispersion theory. The number of "dispersion electrons" one found by fitting the dispersion formula to the experimental data was much lower than one would expect. Drawing on Einstein's quantum theory of radiation to replace amplitudes of radiation by probabilities of emission or absorption, Ladenburg replaced numbers of electrons by numbers of electron jumps and thus arrived at a formula that at least to some extent takes care of this problem. Ladenburg had no solution, however, for the problem that dispersion appears to be anomalous at the wrong frequencies in the old quantum theory. He just replaced orbital frequencies with transition frequencies in the classical formula without any theoretical justification because this was clearly what the experimental evidence indicated. Another limitation of Ladenburg's formula was that it only applied to the ground state of an atom.[14]

---

12. On dispersion in classical theory and the old quantum theory, see Jordi Taltavull (2017).

13. The term "correspondence principle" does not occur in the main body of Bohr (1918, Parts I and II). It does appear, however, in an appendix to Part III, which finally saw the light of day in November 1922, although a manuscript existed already in 1918 as the first two parts went to press. For further discussion of the correspondence principle, see, e.g., Darrigol (1992), Fedak and Prentis (2002), Bokulich (2008), Rynasiewicz (2015), and Jähnert (2019).

Ladenburg's work drew the attention of Bohr and, in 1924, his assistant Kramers found a generalization of Ladenburg's formula that removed its limitations. Kramers combined the sophisticated techniques the old quantum theory had borrowed from celestial mechanics with Bohr's correspondence-principle approach. He considered the scattering of an electromagnetic wave by some generic mechanical system with one electron and derived a classical dispersion formula for such a system that has the form of a derivative with respect to an action variable of an expression containing amplitudes and frequencies of the oscillations induced in that system by an electromagnetic wave. To turn this into a quantum formula, Kramers replaced amplitudes by transition probabilities (as Ladenburg had done before him), orbital frequencies by transition frequencies, and—and this was Kramers's main innovation—derivatives by difference quotients. The construction of this quantum formula guaranteed that it merges with classical theory in the limit of high quantum numbers. In this limit, after all, transition frequencies and orbital frequencies can be used interchangeably and the difference between successive integers in the allowed values of the action variable (which, as Schwarzschild had first shown, was just Sommerfeld's phase integral) becomes so small that derivatives can be replaced by difference quotients. In the spirit of Bohr's correspondence principle, Kramers now took the leap of faith that this formula would continue to hold all the way down to the lowest quantum numbers.[15]

Kramers initially only published his formula in two short notes in *Nature*. Over the Christmas break of 1924–1925, however, he teamed up with Heisenberg, a former student of Sommerfeld's who was visiting Copenhagen, to write a paper providing a detailed exposition and further extension of the results he had found before. This paper played a central role in the train of thought that led to the famous *Umdeutung* (reinterpretation) paper with which Heisenberg laid the foundation for matrix mechanics. One of the striking features of the Kramers dispersion formula is that it only depends on transitions between the orbits used in its derivation. It no longer refers to individual orbits. This seems to have given Heisenberg the key idea of setting up a new framework for all of physics in which any quantity that used to be represented by a number connected to one particular orbit is represented instead by an array of numbers connected to all possible transitions between orbits. The reason Heisenberg referred to this as *Umdeutung* is that he retained the laws of classical mechanics relating these quantities. He only reinterpreted the nature of the quantities related by these laws. Heisenberg emphasized that all observable quantities (frequencies, intensities, and polarizations of radiation) correspond not to individual orbits but to transitions between them. He hoped to eliminate the increasingly problematic orbits altogether by focusing on transitions between stationary states.[16]

---

14. Ladenburg (1921), Einstein (1917).

15. Kramers (1924a, 1924b). Full derivations of the Kramers dispersion formula were first published by Born (1924) and Van Vleck (1924a, 1924b).

To achieve this goal, Heisenberg also needed to replace the basic Bohr-Sommerfeld quantization condition, which, after all, gave the allowed values of the action variable for individual orbits. In keeping with his *Umdeutung* program, Heisenberg looked at the change in values of these action variables in transitions between orbits. Transcribing the classical formula for such changes into a quantum one, in a manner analogous to what Kramers had done in dispersion theory, Heisenberg arrived at a formula he had encountered before. Right around that time and independently of one another, Werner Kuhn in Copenhagen and Willy Thomas in Breslau had derived an expression for the high-frequency limit of the Kramers dispersion formula that has come to be known as the Thomas-(Reiche-)Kuhn sum rule. This is what Heisenberg used as his quantization condition in the *Umdeutung* paper.[17]

Heisenberg had not just studied with Sommerfeld in Munich and spent time in Bohr's institute in Copenhagen, he had also co-authored a paper with Born in Göttingen. In the early 1920s and under Born's leadership, Göttingen had emerged alongside Copenhagen and Munich as a third leading center for work on the old quantum theory. When Born read the *Umdeutung* paper, he immediately recognized that the arrays of quantities Heisenberg had introduced were nothing but matrices and that Heisenberg's peculiar non-commutative multiplication was nothing but the standard rule for matrix multiplication. He also realized that Heisenberg's quantization condition, the Thomas-Kuhn rule, is equivalent to the diagonal elements of the commutation relation $\hat{q}\,\hat{p}-\hat{p}\,\hat{q} = i\,\hbar$ (where hats indicate that these quantities are matrices). His student, Jordan, showed that the off-diagonal elements vanish. They reported these results in a joint paper elaborating on Heisenberg's *Umdeutung* paper. Heisenberg generalized this commutation relation from one to multiple degrees of freedom. The resulting commutation relations, $[\hat{q}_k, \hat{p}_l] = i\,\hbar\,\delta_{kl}$, are central to the first authoritative exposition of matrix mechanics, the *Dreimännerarbeit* of Born, Heisenberg, and Jordan.[18]

Around the same time and independently of the work of Born and Jor-dan, Dirac, using arguments from dispersion theory, derived a precise correspondence between classical Poisson brackets and the commutator of the corresponding quantum variables. Dirac's procedure precisely imitates the methodology of the Kramers-Heisenberg derivation of the dispersion formula for inelastic (Raman) light scattering, in which the amplitude for a transition between two quantum states $a$ and $b$ was expressed as a sum of amplitudes for transitions via two distinct intermediate states $c$ and $d$ (i.e., as a sum of the amplitudes for the sequential transitions $a \to c \to b$ and $a \to d \to b$). The transition between classical formulas involving derivatives with respect to the action and quantum ones involving discrete differences of amplitudes with varying quantum numbers was accomplished via the transcription procedure $d/dJ \to 1/h\, \Delta/\Delta n$ by now familiar from the work of Kramers, Born, Van Vleck, and Heisenberg. Dirac's beautiful derivation shows that the commutator of any two kinematical variables (divided by Planck's constant $h$), interpreted according to Heisenberg's matrix reinterpretation, corresponds precisely to the Poisson bracket of the associated classical quantities in the limit of large quantum numbers.[19]

---

16. Kramers and Heisenberg (1925). On the path leading from Kramers's dispersion theory to Heisenberg's (1925) *Umdeutung* paper, see, e.g., Dresden (1987) and Duncan and Janssen (2007).
17. Kuhn (1925), Thomas (1925), Reiche and Thomas (1925).
18. Born and Heisenberg (1923), Born and Jordan (1925), Born, Heisenberg, and Jordan (1926).

Completely independently of the work of the Göttingen group, and of Dirac in Cambridge, a formulation of quantum theory based on a continuum wave theory was developed in late 1925 and early 1926 by Schrödinger, working in Zurich. The theory, inspired by the work of Louis de Broglie on matter waves (via Einstein's second paper on the quantum theory of the ideal gas), and by the analogies already discovered almost a century earlier by William Rowan Hamilton between geometrical optics and particle mechanics, posited the existence of a well-defined solution $\psi(\vec{r})$ to a wave equation associated in some way with the dynamics of a single particle. In his first paper on wave mechanics published in January 1926, this wave equation was obtained as the solution of a variational problem based on the classical Hamilton-Jacobi equation, and the appearance of energy quantization—for the bound states, with negative energy, of the hydrogen atom—is a consequence of the imposition of regularity and finiteness conditions on the wave function $\psi$. Once the wave function $\psi(\vec{r})$ is required to *remain finite* as $r \to \infty$, the Bohr-Balmer quantization of the bound states of the hydrogen atom follows immediately. A few months later, Schrödinger (and independently, Wolfgang Pauli in Hamburg and Carl Eckart at Caltech) established the connection between his wave functions and the matrices of Heisenberg *et al.*, at which point it became clear that the matrices so defined would also satisfy the commutation relation which served as the point of departure of matrix mechanics.[20]

---

19. Dirac (1926a, 1926b). For analysis, see Darrigol (1992) and Kragh (1990).

20. De Broglie (1924, 1925), Einstein (1925), Schrödinger (1926a, 1926b), Eckart (1926), Pauli to Jordan, April 12, 1926 (Pauli 1979, pp. 315–320; translated and discussed by van der Waerden 1973). For discussion of Schrödinger's use of the optical-mechanical analogy, see Joas and Lehner (2009); for discussion of his equivalence proof, see Muller (1997–1999).

In late 1926, Jordan and Dirac independently of one another found essentially the same formalism, now known as the Dirac-Jordan statistical transformation theory, unifying matrix mechanics, wave mechanics, Dirac's $q$-number theory and yet another version of the new quantum theory, the operator calculus of Born and Norbert Wiener. A few months later, drawing on Jordan's work, Heisenberg showed that the commutation relations central to the new theory express what we now know as the uncertainty principle. Around the same time, John von Neumann introduced the Hilbert space formalism of quantum mechanics and showed that matrix mechanics and wave mechanics correspond to two different incarnations of Hilbert space, the space $l^2$ of square-summable sequences and the space $L^2$ of square-integrable functions, respectively.[21]

## 5 Conclusion

As this brief overview shows, the canonical commutation relations $\hat{q}_i\,\hat{p}_j - \hat{p}_j\,\hat{q}_i = i\,\hbar\delta_{ij}$ at the heart of modern quantum mechanics can be traced back to Heisenberg's use of the Thomas-Kuhn sum rule, a corollary of the Kramers dispersion formula, as the quantization condition in his *Umdeutung* paper. Heisenberg was led to this quantization condition by transcribing (in a manner analogous to how Kramers arrived at his dispersion formula) the phase-integral quantization condition $\oint p\,dq = nh$ of the old quantum theory found by Sommerfeld, Wilson, Ishiwara and clarified by Schwarzschild and Epstein, who identified and exploited the connection of these phase integrals to the action variables familiar from celestial mechanics. What, in turn, had inspired Sommerfeld to adopt the phase-integral quantization condition was Planck's reworking of the condition $\varepsilon = h\nu$ central to the derivation of his blackbody radiation law.

## 6 Acknowledgment

We thank Olivier Darrigol, Olival Freire Jr., and Alexei Kojevnikov for helpful comments on an earlier draft of this essay.

---

21. Born and Wiener (1926), Jordan (1927a, 1927b), Dirac (1927), Heisenberg (1927), von Neumann (1927). For discussion of the transition from Jordan's transformation theory to von Neumann's Hilbert space formalism, see Duncan and Janssen (2013).